\documentclass[doublecol]{epl3}
\usepackage{amsmath}

\newtheorem{lemma}{Lemma}
\newtheorem{defi}{Definition}

\title{Survival of entanglement in thermal states}
\shorttitle{Survival of entanglement in thermal states} 

\author{Damian Markham\inst{1} \and Janet Anders\inst{2} \and Vlatko Vedral\inst{2,3} \and Mio Murao\inst{1,4} \and Akimasa Miyake\inst{5}}
\shortauthor{D. Markham \etal}

\institute{
  \inst{1} Department of Physics, Graduate School of Science,
  University of Tokyo, Tokyo 113-0033, Japan\\
  \inst{2} Quantum Information Technology Lab, Department of
Physics, National University of Singapore, Singapore 117542, Singapore\\
  \inst{3} The School of Physics and Astronomy, University of
Leeds, Leeds LS2 9JT, UK\\
  \inst{4} PRESTO, JST, Kawaguchi, Saitama 332-0012, Japan\\
\inst{5} Institute for Theoretical Physics, University of Innsbruck, Technikerstra{\ss}e 25, A-6020 Innsbruck, Austria \\ Institute for Quantum Optics and Quantum Information, Austrian Academy of Sciences, Innsbruck, Austria.
}


\abstract{We present a general sufficiency condition for the presence of multipartite entanglement in
thermal states stemming from the ground state entanglement. The condition is written in terms of the ground state entanglement and the partition function and it gives transition temperatures
below which entanglement is guaranteed to survive. It is flexible and can be easily adapted to consider
entanglement for different splittings, as well as be weakened to allow easier calculations by
approximations. Examples where the condition is calculated are given. These examples allow us to
characterize a minimum gapping behavior for the survival of entanglement in the thermodynamic limit. Further, the same technique can be used to find noise thresholds in the generation of useful resource states for one-way quantum computing.}

\begin{document}

\maketitle

In recent years there has been much effort to investigate the role of entanglement in general physics
problems. Entanglement is known to be a key resource for quantum information, essential for faithful
teleportation and allowing an absolute secure key distribution among other things \cite{NielsenChuang},
and the study of entanglement has been mainly from this perspective. However, entanglement is also a
key foundational issue in quantum mechanics, and has recently been associated to various phenomena in
different areas of physics, for example Hawking radiation in cosmology \cite{Callan94}, symmetry
breaking in high energy physics \cite{Bertlmann01} and in particular, to many areas of condensed matter
physics such as critical phenomena \cite{Osborne02}. In addition, entanglement theory has also been
helpful in finding the ground state for difficult many-body systems \cite{Vidal04}.

All of these results and connections are very intriguing, and lead us to ask when and where else
entanglement exists, and what is its role in the associated phenomena. To address these issues, in the
first instance, it would be very useful to have a general, easy test to see if a system contains
entanglement. There are several difficulties in this, especially in many-body physics. First, in
nature, systems are in thermal states, and calculating the density matrix involves the diagonalisation
of large Hamiltonians which, in general, proves impossible. Second, given this state density matrix, it
is difficult to test if it is entangled or not.

One situation that helps simplify the problem is given by systems where the ground state is highly
entangled.  This is often the case for symmetric or interacting many-body systems. There, at low enough
temperatures, the properties are governed by the ground state, and the system is entangled. Along these
lines, for example, in \cite{Brukner04,06Anders}, by taking the minimum expectation of the energy
allowed for separable states, conditions for entanglement are found on the average energy and
associated thermodynamic quantities, so observing energy below this minimum value means the system is
entangled.

In this paper we present an explicit connection between the ground state entanglement and the
entanglement of the thermal state. We give a condition for the existence of entanglement based on
minimum knowledge of the ground state and statistical properties of the full state (the partition
function), so that we do not need to calculate the full density matrix. This is a kind of coarse grain
approach to the existence of entanglement. Weaker, more easily calculable versions using approximations
to these quantities follow easily. The methods can further be adapted to consider different types of
entanglement (from the full multipartite entanglement to bipartite entanglement), and also give ever
more precise entanglement conditions, up to the exact case. We present several examples which include
extreme (good and bad) cases, allowing us to observe a minimum requirement on the gapping behaviour of the Hamiltonian if
we are to see entanglement from the ground state survive in the thermodynamic limit.

%
%

\bigskip

Consider a general system in the thermal state
\begin{equation} \label{eq:thermal}
    \rho_T = e^{- H / k_B T}/Z,
\end{equation}
where $H$ is the Hamiltonian, $T$ is the temperature, $Z$ is the partition function and  $k_B$ is the Boltzmann constant. The thermal states can be
expanded in their diagonal basis, the energy eigenbasis, as
\begin{eqnarray}\label{eq:expandthermal}
    \rho_T = {e^{- E_0 /k_B T} \over Z} |e_0 \rangle \langle e_0 | +
    \sum_{j=1} {e^{- E_j /k_B T} \over Z} |e_j \rangle \langle e_j |,
\end{eqnarray}
where $|e_0\rangle$ is the lowest energy eigenstate, which is often entangled and  $|e_j \rangle$,
$j>0$, are the (possibly entangled) excited states having energies $E_j > E_0$. For any temperature $T
\not = 0$ the thermal state is mixed except for $T= 0$ when $\rho_T$ becomes identical to the pure
ground state (the lowest energy eigenstate $|e_0\rangle$). It is apparent that if the lowest energy
eigenstate, $| e_0 \rangle$, is entangled, then the system will be entangled at $T=0$ and it seems
reasonable that it will retain at least partial entanglement for small, finite temperatures. The
immediate question is, when $T$ grows, and a portion of excited states gets mixed in, is the thermal
state for such finite $T$ still entangled? And what is the critical temperature for the survival of
that ground state entanglement?  To answer this question we pick an entanglement measure that suits the
problem best: The {\it global robustness of entanglement} \cite{Vidal99} measures the worst degradation
of entanglement through mixing and allows therefore to establish a minimal temperature range in which
the ground state entanglement survives within the thermal mixture.

\begin{defi} \label{def:robustness}
The global robustness of entanglement, $R(\sigma)$, is defined for a state, $\sigma$, as the minimum
amount, $t$, of arbitrary noise, $\tau$, that needs to be mixed to $\sigma$ to make the state
separable, i.e.
\begin{equation} \label{eq:DEFrobustness1}
    R(\sigma) := \min_{\omega(\tau, t)} t,
\end{equation}
such that there exists a state $\tau$ satisfying
\begin{equation} \label{eq:DEFrobustness2}
    \omega (\tau, t) := {1 \over 1+t }(\sigma + t\tau)\in \cal{S},
\end{equation}
where $\cal{S}$ is the set of separable states with respect to an arbitrary partition of the system
into subsystems. If $\sigma$ is separable with respect to that partitioning, then no noise needs to be
added and the robustness is zero, $R(\sigma)=0$. Conversely,  $R(\sigma) > 0$ if and only if $\sigma$
is entangled.
\end{defi}

\begin{figure}[t]
    \begin{center}
    \resizebox{!}{3.6cm}{\includegraphics{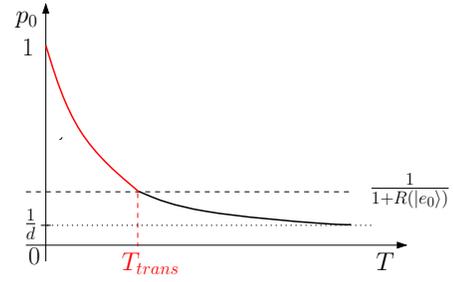}}
    \caption{\label{fig:p0} The probability of the lowest energy eigenstate, $p_0 ={e^{- E_0 /k_B T} \over Z }$, is plotted against temperature $T$ for a typical thermal state. The point where the curve crosses the value ${1 \over 1 +R(|e_0 \rangle)}$ defines the transition temperature $T_{trans}$, below which entanglement is guaranteed.}
    \end{center}
\end{figure}

Taking a thermal state $\rho_T$, expanded as in Eq.~(\ref{eq:expandthermal}), as a candidate for
$\omega$  in Eq.~(\ref{eq:DEFrobustness2}) one can infer how much thermal add-mixing is at least
allowed until all entanglement could vanish. This can be achieved by identifying, for instance, the
lowest energy eigenstate $|e_0\rangle$ with the state $\sigma$ and the higher energy eigenstates as
thermal noise. In this scenario the pre-factor of $\sigma$ is just the probability of the lowest energy
eigenstate within the thermal mixture, i.e. $ {1 \over 1 +t} ={e^{- E_0 /k_B T} \over Z}$. This
argument leads immediately to a sufficient criterion for the presence of entanglement at finite
temperature, which is stated in Lemma \ref{lem:survival} and illustrated in Fig.~\ref{fig:p0}.

\begin{lemma} \label{lem:survival}
A thermal state $\rho_T$ as defined in Eq.~\eqref{eq:thermal}, with partition function, $Z$, and whose
lowest energy eigenstate, $|e_0\rangle$, has energy $E_0$ and global robustness of entanglement
$R(|e_0)\rangle$, must be entangled if
\begin{equation} \label{eq:survival}
    {e^{- E_0 /k_B T} \over Z}  > {1 \over 1 +R(|e_0 \rangle)}
\end{equation}
holds.
\end{lemma}

The Lemma follows directly from the definition of the global robustness of entanglement and is not
surprising with respect to the mathematics. However, it is worth discussing the physical aspects of the
use of this adapted measure for thermal states. In particular it is interesting to note that relation
\eqref{eq:survival} combines some statistical properties of the thermal state, such as the partition
function and the ground state energy, with the entanglement content of the ground state. The fulfilment
of condition \eqref{eq:survival}  detects or \emph{witnesses} the existence of entanglement in the
thermal mixture, even though entanglement can also exist when Eq.~\eqref{eq:survival} is not met.
Despite this fuzzy transition, it is interesting to define a \textit{transition temperature}
$T_{trans}$ by setting equality in \eqref{eq:survival}, see Fig.~\ref{fig:p0}, which tells when
entanglement must necessarily occur in the system. To make a first guess whether a thermal state is
entangled and give a minimal transition temperature only the three ingredients stated above are needed
- as opposed to the full knowledge of all energy eigenstates or the entanglement properties of higher
excited energy states.  Of course, knowing these could lead to more accurate bounds on the temperature
range where entanglement is present. The condition itself however is only tight, when the thermal
mixture degrades the entanglement of the ground state in the most efficient way. Indeed entanglement of the thermal state may go up as the temperature increases, however, this would not be due to the ground state's entanglement, and it is typically not so, except in very special examples.
\bigskip

{\bf Obervations}: Let us make a few observations about the validity and power of condition \eqref{eq:survival}. Firstly,
it can be simply adapted to also consider the situations where the main entanglement contribution does
not come from the ground state, but some excited state $|e_j\rangle$, or even contributions from
several states. This is done by replacing $R(|e_0\rangle)$ with the robustness of the state in
question, $R(|e_j\rangle)$, in the right-hand-side (RHS) of Eq.~\eqref{eq:survival} and the state's
population $p_j = {e^{- E_j /k_B T} \over Z}$ on the left-hand-side (LHS). Similarly we can adopt
degenerate ground states by replacing the state $|e_0\rangle$ by the maximally mixed state across the
degenerate space. We could also, for example, take the first few excited states as the main
contributor, right up to considering the whole state i.e. including all high energy levels. Of course,
the price to pay is that the robustness of entanglement will become more difficult or even impossible
to calculate.

Secondly, by considering the entanglement of the ground state with respect to different partitions into
subsystems the condition can witness different types of entanglement. A system of many parties can be
re-arranged by grouping some of the parties together, for instance as a bipartite grouping, and the
global robustness of entanglement for such grouping could be applied in Eq.~\eqref{eq:survival} as well
as the robustness of the full multi-partite partitioned system. Of course the set of  totally separable
states with respect to all parties is a subset of the set of bipartite separable states and therefore
the robustness decreases the fewer parties we distinguish, $R(\rho)\geq R_{Bi}(\rho)$. The LHS of
Eq.~\eqref{eq:survival} is a monotonically decreasing function of temperature so that higher ground
state entanglement implies a higher transition temperature.  Therefore, the corresponding transition
temperature for a bipartite grouping, $T_{trans}^{Bi}$, must be less or equal then the transition
temperature for full multi-partite entanglement, $T_{trans}^{Bi} \leq T_{trans}$. This is
similar to the situation in \cite{06Anders}, where the transition temperature increases with the number
of partitions. Fortunately, $R_{Bi}(\rho)$ is known for pure states \cite{Vidal99} (for any bipartite
cut), so it is always possible to calculate a bound to the RHS of Eq.~\eqref{eq:survival} by
considering only bipartite cuts for a given ground state.

\begin{figure}[h]
    \begin{center}
    \resizebox{!}{3.0cm}{\includegraphics{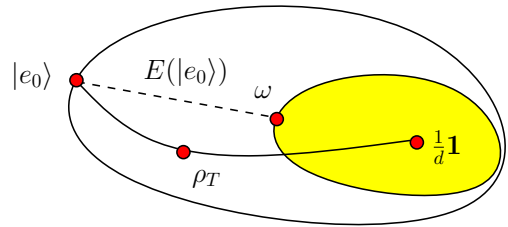}}
        \caption{\label{fig:statespace} The figure shows the set of all
states and the sub-set of separable states (shaded area). The distance between the ground state
$|e_0\rangle$ and the closest separable state $\omega$ gives the entanglement of the ground state
$E(|e_0\rangle)$, illustrated by the dotted line. The solid line is the  line of thermal states
$\rho_T$ for increasing temperature from $T=0$ ($\rho_0 = |e_0\rangle \langle e_0|$) to $\rho_{\infty}
= {1 \over d} {\bf 1}$ for $T \to \infty$. If the distance from $|e_0\rangle$ to $\rho_T$ is smaller
than distance from $|e_0\rangle$ to $\omega$, then $\rho_T$ must be entangled.}
    \end{center}
\end{figure}

Thirdly observe, that when the partition function and the robustness of entanglement is replaced by
upper and lower bounds to those quantities, respectively, Eq.~\eqref{eq:survival} is still satisfied
and presents a valid entanglement condition. Thus, we may use approximation methods for all
calculations, providing they bound in the correct direction. An example of such a lower bound to the
global robustness of entanglement is given in \cite{Hayashi05}. For any pure state, $| \psi \rangle$,
\begin{eqnarray} \label{eq:Ent-inequality}
    1+ R(|\psi\rangle) \geq 2^{E_R(|\psi\rangle)} \geq
    2^{E_G(|\psi\rangle)},
\end{eqnarray}
where $E_R(|\psi\rangle)$ is the {\it relative entropy of entanglement} \cite{Vedral98} and
$E_G(|\psi\rangle)$ is the {\it geometric measure of entanglement} \cite{Wei03}. These distance-like
entanglement measures allow an intuitive interpretation of the entanglement condition
\eqref{eq:survival}. One can understand $E_R(|\psi\rangle)$ and $E_G(|\psi\rangle)$ as the minimum
``distance'' $D(|\psi\rangle||\omega)$ to the closest separable state $\omega$ with respect to relative
entropy and the geometric overlap respectively. Thus, if $D(|e_0\rangle||\rho_T)<D(|e_0\rangle||\omega)
= E(|e_0\rangle)$ then $\rho_T$ is entangled (illustrated in Fig.~\ref{fig:statespace}).

For the thermal state, both the relative entropy and the geometric measure can be easily calculated
giving $D(|e_0\rangle||\rho_T)=-\log(p_0)$, where $p_0=e^{-E_0 /k_B T} / Z$ is the ground state
population. This relation reproduces \eqref{eq:survival},
\begin{eqnarray} \label{eqn: condition statistical}
    p_0 > 2^{-D(|e_0\rangle||\omega)},
\end{eqnarray}
when $R(|e_0\rangle)$ is replaced with the bounds in \eqref{eq:Ent-inequality}. The statistical
interpretation of the relative entropy \cite{Vedral97b} on the RHS of (\ref{eqn: condition
statistical}) offers another, statistical, interpretation of Eq.~\eqref{eq:survival} and
Eq.~\eqref{eqn: condition statistical}: Entanglement is present if the probability of the system being
in the ground state is greater than the (asymptotic) probability that $|e_0\rangle$ would be mistaken
for its closest separable state $\omega$.


\bigskip

{\bf Example 1:} We now consider three illustrative examples for which we derive transition temperatures for entanglement.
Our first example is a spin dimer, that is, two spin-$1/2$ particles interacting via  the Heisenberg
Hamiltonian with coupling $J$, exposed to an external magnetic field $B$,
\begin{eqnarray}
H=B(\sigma_Z^1+\sigma_Z^2) + J(\sigma_X^1\sigma_X^2+\sigma_Y^1\sigma_Y^2+\sigma_Z^1\sigma_Z^2),
\end{eqnarray}
where $\sigma^i_{X/Y/Z}$ are the Pauli matrices for the $i$th spin and $B,J\geq 0$. This example allows
us to compare our condition \eqref{eq:survival} to the exact case which was calculated in
\cite{Arnesen01}.

Diagonalising the Hamiltonian leads to the four eigenvalues $J+B, J, J - B, -3 J$ from which we can
immediately calculate the LHS of \eqref{eq:survival}. We have $Z =
e^{-(J+B)/k_BT}+e^{-J/k_BT}+e^{-(J-B)/k_BT}+e^{3J/k_BT}$ and for $B < 4J$ the ground state is the
singlet with eigenvalue $-3J$. The robustness of the singlet is known, $R(|e_0\rangle)=1$
\cite{Vidal99}.  So, for $B < 4J $ we have the sufficient condition for entanglement
\begin{eqnarray} \label{eqn: Ent Condition Dimer}
e^{-4J/k_BT}(e^{B/k_BT}+e^{-B/k_BT}+1) < 1.
\end{eqnarray}
In particular, when $B =0$, the excited states are all degenerate with eigenvalue $J$, and  the
entanglement  condition can be rewritten to give a transition temperature below which the system is
guaranteed to be entangled
\begin{eqnarray}
T<\frac{4J}{k_B\ln{3}} \equiv T_{trans}.
\end{eqnarray}
This remarkably coincides with the exact entanglement result in \cite{Arnesen01}. For non-zero $B$ the
excited states lose their degeneracy and bound (\ref{eqn: Ent Condition Dimer}) gives a lower
transition temperature than the exact result. This happens because our inherent approximation that the
thermal mixture kills off the entanglement optimally, becomes incorrect.

When $B > 4J$, the singlet ceases to be the ground state, and is instead superseded by the separable
state $|00\rangle$ with energy $J - B$. The system thus undergoes a quantum phase transition.
Expression (\ref{eqn: Ent Condition Dimer}) still gives a valid sufficient condition for entanglement,
but now the main contribution comes from the first excited state (in this case the singlet), rather
than the ground state (which is no longer entangled). However it is no longer satisfied for any
temperature, and so does not see the presence of entanglement, even though we know it exists from
\cite{Arnesen01}.


\bigskip

{\bf Example 2:} The second example is a simplistic very general model for a class of many-body Hamiltonians, which
allows us to examine extreme cases, and make interesting general statements. We only specify the energy
spectrum (giving the LHS of \eqref{eq:survival})  and the ground state, common in many-body physics (whose robustness gives us the RHS of \eqref{eq:survival}). This toy model illustrates the
flexibility we have in considering these quantities separately.

The ground states we consider are of the form
\begin{eqnarray} \label{eqn: DEF symm state}
    |S(n,k)\rangle := \frac{1}{\sqrt{{n\choose k}}}
    (\sum_{PERM}|\underbrace{00...0}_{n-k}\underbrace{11..1}_k\rangle),
\end{eqnarray}
where the sum is over all permutations of position. These states describe $n$ systems with $k$
excitations symmetrised over all permutations. We could, for example, imagine  $n$ sites, which can
either be empty or occupied by
 $k$ identical (fermion) particles. These are
typical ground states in solid state physics, for example in some Hubbard and related models. They also
arise as Dicke states when $n$ atoms interact with a single electromagnetic mode \cite{Vedral04}.

It can be shown \cite{Hayashi07} that the symmetry of these states means that the global robustness of
entanglement is related to the relative entropy of entanglement ($E_R$) as
\begin{eqnarray}
1+R(|S(n,k)\rangle) = 2^{E_R(|S(n,k)\rangle)}.
\end{eqnarray}
The relative entropy of entanglement for the states $|S(n,k)\rangle$ is known  \cite{Vedral04,Wei04}
and gives
\begin{eqnarray}
1+R(|S(n,k)\rangle)= {1 \over {n\choose k}}\left( {n \over k} \right)^k \left( {n \over n-k}
\right)^{n-k}.
\end{eqnarray}
For simplicity we now only consider the case where $k=n/2$. This choice of $k$ maximises the
entanglement and for large $n$ we have
\begin{eqnarray} \label{eqn: Ent s(n,n/2)}
1+R(|S(n,n/2)\rangle) = 2^{E_R(|S(n,k)\rangle)}= \sqrt{n}.
\end{eqnarray}

For the LHS of \eqref{eq:survival} we consider Hamiltonians of the form
\begin{eqnarray}
H = E_0 | e_0 \rangle \langle e_0| + \sum_{m=1}^{D-1} (E_0 + m^{\alpha} \Delta) |e_m \rangle\langle
e_m|
\end{eqnarray}
where $D$ is the number of energy levels (the dimension of the space), with spacing parameter $0 \le
\alpha \le 1$ and $\Delta$ is the energy gap between the ground state and the first excited state. This gives a range of energy spectra from the case where all excited states are
degenerate ($\alpha=0$) to the case where they are equally spaced ($\alpha=1$). Although this model is
rather simplified and hence somewhat artificial, as mentioned, its simplicity allows us to examine
extreme cases, through which we are able to make general statements of interest true for all systems.

For simplicity here, we will use the relative entropy of entanglement to consider transition
temperatures. As in the above entanglement calculation, it will not effect any of our conclusions to
use this weaker condition.

The thermodynamical partition function is
\begin{eqnarray} \label{eqn: Z_alpha}
Z_\alpha =e^{- E_0/k_BT}\left(1+ \sum_{m=1}^{D-1}e^{- m^{\alpha} \Delta/k_BT}\right).
\end{eqnarray}
For $\alpha=0$, the degenerate case, we can easily evaluate the sum in (\ref{eqn: Z_alpha}), giving the
condition for entanglement
\begin{eqnarray} \label{eqn: T_0}
T<\frac{\Delta}{k_B}\frac{1}{\ln\left({\frac{D-1}{2^{E_R(|e_o\rangle)}-1}}\right)}\equiv T_0.
\end{eqnarray}
In fact $Z_0$ is an upper bound to any partition function. Hence $T_0$ in (\ref{eqn: T_0}) is a lower bound to \textit{all} possible threshold temperatures, written only in terms of the gap to the first excited state $\Delta$ and the entanglement of the ground state.

However, in the thermodynamic limit, this lower bound will almost certainly tend to zero. We can see this by noting that the entanglement of any state is trivially upper bounded by $\log{D}$. This is
found by considering the distance to the maximally mixed state. In general this value cannot be
achieved \cite{Zyczowski98}, and so $E< \log{D}$, and at best we can hope for the dimension over entanglement term to give a constant in scaling. More usual (in fact as far as the authors know, there are no counter examples), will be as in the example state (\ref{eqn: Ent s(n,n/2)}) where the entanglement is so low that this term will cause the temperature to tend to zero in the limit.

Of course we must also consider the scaling of $\Delta$, but again, at best this is likely to be
constant, and at worst (and more usually) it will also limit to zero in the thermodynamic limit. We can
conclude that in this $\alpha=0$ ``worst case'' spectrum, even with a highly entangled ground state, the entanglement cannot survive any non-zero temperature in the thermodynamical limit.

For $\alpha=1$ we get $Z_1=e^{-E_0/k_BT}\sum_{m=0}^{D-1}e^{- m\Delta/k_BT}$. In the limit of
$D\rightarrow \infty$ we find the entanglement condition
\begin{eqnarray}
T<
\frac{\Delta}{k_B}\frac{1}{\ln\left({\frac{2^{E_R(|e_o\rangle)}}{2^{E_R(|e_o\rangle)}-1}}\right)}\equiv
T_1.
\end{eqnarray}
When we take a low temperature approximation of $Z_1$ the transition temperature $T_1$ scales roughly
with $2^{E_R(|e_o\rangle)}$
\begin{eqnarray} \label{eqn: lowt1} T_1 \approx
\frac{\Delta}{k_B}2^{E_R(|e_o\rangle)}.
\end{eqnarray}
So, in the limit of large $n$, for fixed $\Delta$,  we have entanglement for arbitrarily high
temperatures. This spectrum is in some sense a best case extreme, and we would not expect such a nice
situation in real systems. However it does allow us to see that if we wish to see entanglement in
the thermodynamical limit, the scaling of $\Delta$ should not be too bad compared to the scaling of the
entanglement.

Thus, to see entanglement, even for the best case energy spectrum, we require an energy
gap between the ground and first exited state that scales at least of the order
\begin{eqnarray} \label{eqn: gapping rule}
\Delta \geq \frac{1}{2^{E_R(|e_o\rangle)}}.
\end{eqnarray}
Any less would mean that the transition temperature tends to zero in the thermodynamic limit.

If we take the example state entanglement (\ref{eqn: Ent s(n,n/2)}), we see the transition temperature
\begin{eqnarray}
\label{eqn: lowt2} T_1 \approx \frac{\Delta}{k_B}\sqrt{n}.
\end{eqnarray}
Now, the way $\Delta$ scales with $n$ is very dependant on the system. To see entanglement we would
want a system with scaling $\Delta \geq 1/\sqrt{n}$.

We now consider the intermediate case for completeness. For $\alpha \neq 0$ and in the limit
$D\rightarrow \infty$ we can approximate the sum in (\ref{eqn: Z_alpha}) by an integral  which has the
same form as Riemann's gamma function $\Gamma$  giving
\begin{eqnarray} \label{eqn: Z alpha approximation}
Z_\alpha \approx e^{-E_0/k_BT}
\frac{\Gamma\left(\frac{1}{\alpha}\right)}{\alpha}\left(\frac{k_BT}{\Delta}\right)^{1/\alpha}.
\end{eqnarray}
Thus, with (\ref{eqn: Ent s(n,n/2)}) we get the condition for entanglement
\begin{eqnarray} \label{eqn: CritTempAlpha1}
T<\frac{\Delta}{k_B} \left[ \frac{\alpha \sqrt{n}}{\Gamma\left( \frac{1}{\alpha}\right)}
\right]^\alpha\equiv T_\alpha.
\end{eqnarray}
For $\alpha=1$ this reduces to the low temperature approximation (\ref{eqn: lowt2}). Similarly, for all
$\alpha>0$ the transition temperature increases with $n$. Indeed for any ground state with entanglement scaling with $n$ a similar statement holds. Again we see, that for our technique to see
the survival of entanglement from the ground state, we would need some limits on the scaling of the gap $\Delta$. This is in analogy to many situations in condensed matter physics, where to witness interesting phenomena we need a gapped system.

As mentioned these examples are somewhat artificial, and realistic Hamiltonians will lie somewhere in between the $\alpha=0$ and $\alpha=1$ cases. However, with the fact that the gap plays such an important role in both extremes (and for in between $\alpha$), and the fact that as soon as we step out of the $\alpha=0$ case we have non-zero threshold temperature, even when the gap scales to zero in the limit (albeit conditioned that it does not do so too fast), we may conjecture that for any realistic Hamilitonian, with fixed gap $\Delta$, any scaling of the entanglement of the ground state with $D$ implies a non-zero threshold temperature.

\bigskip
{\bf Example 3:} Our third and final example is that of the so called stabiliser Hamiltonian, whose ground state is an important resource for quantum information, most notably for one-way quantum computing \cite{Briegel01}.
\begin{eqnarray} \label{eqn: stabiliser hamiltonian}
H=-B \sum_{i=1}^{n}K_i, \; \; \; K_i:= \sigma_x^i \otimes_{j\in N(i)} \sigma_z^j,
\end{eqnarray}
where the $K_i$ are the so called generators of the stabilizer group, and $N(i)$ are the neighbours of $i$ in a given lattice (or graph). Although this Hamiltonian is not a naturally occuring one in many-body physics, there are proposals for its implementation. Further, its thermal state is also identical to the mixed state induced by several realistic noise models for the generation of cluster states in various schemes, for example using optical lattices, (see e.g. \cite{Kay06}).

The ground state of this Hamiltonian is the graph state of the associated lattice (or graph). The excited states are achieved by making local $\sigma_z^i$ flips, acting as an excitation operator. For all lattices, the ground state has energy $-nB$, and the $i$th excited state has energy $B(-n+2i)$ with degeneracy $ n \choose i$. The partition function is easily found as
\begin{eqnarray}
 Z=e^{nB/kT}\left(\frac{1+e^{2B/kT}}{e^{2B/kT}} \right)^n.
\end{eqnarray}
Our condition for the survival of entanglement then reads (again taking relative entropy of entanglement)
\begin{eqnarray}
 T< -\frac{2B}{k\ln(2^{E_R(|e_0\rangle)/n}-1)}\equiv T_{trans}.
\end{eqnarray}

This can also be translated into conditions on allowed noise when directly constructing the graph state. For example, suppose a local $\sigma_z^i$ error occurs with equal probability $P$ on any site on such a graph state, this noise state can easily be seen to corresponding to the thermal state of (\ref{eqn: stabiliser hamiltonian}), with $P=\frac{1}{1+e^{2B/kT}}$. Hence this translates to a condition on tolerable flip error threshold, below which the state is still entangled, of
\begin{eqnarray}
 P<1-2^{-E_R(|e_0\rangle)/n} \equiv P_{trans}
\end{eqnarray}

For many graph states, including the 1D, 2D, and 3D cluster states, the entanglement is calculated to be $E_R=n/2$ \cite{Markham06}, giving threshold temperature and noise level of
\begin{eqnarray}
 T_{trans}=-\frac{2B}{k \ln(\sqrt{2}-1)},\;\;\;P_{trans}=1-1/\sqrt{2}.
\end{eqnarray}

Incredibly this coincides exactly with the temperature below which entanglement can be distilled \cite{Kay06}. Even more, in the same paper Kay et al. also show that above this temperature entanglement cannot be distilled. This is very encouraging, since by completely independant methods we have arrived at the same transition values. Thus, here our condition not only implies the existence of entanglement it also turns out that the detected entanglement is of the useful, distillable, kind. Whether this is a general property of the criterion is presently unclear nonetheless an intriguing possibility. Finally, our approach will work for other lattice shapes and perhaps allow extensions where the methods of \cite{Kay06} fail. This is beyond the scope of this work, and we rather leave it for future study.


\medskip

In conclusion, in this paper we have presented a general condition for entanglement in thermal states,
based on the entanglement of the ground state. The condition separates the statistical physics part
from the entanglement part, so that results may be used from either field together to give transition
temperatures below which entanglement is guaranteed. The flexibility of the condition allows for
approximations, allowing easier calculation, and for different kinds of entanglement to be studied.
Further, by considering extreme examples we have seen that even in the best case (in terms of the
energy spectrum, thus $Z$), we need some gapping if entanglement is to survive in the thermodynamic
limit, as characterized by equation (\ref{eqn: gapping rule}). With the final example we see that the approach can be used, not only in many-body physics, but also for simulating noise models for preperation of resource states for quantum information, to find tolerable noise thresholds.

Through $Z$ and the equations of state, the main condition \eqref{eq:survival} can be rewritten in
terms of other thermodynamic quantities as in \cite{06Anders}, allowing the prospect of a variety of
experimentally accessible quantities detecting entanglement. These results should provide a very
useful tool for investigating entanglement in many statistical physics systems.

Several natural questions arise from these investigations. Firstly, with respect to quantum
information, we can ask, given that there are systems which can contain entanglement in the
thermodynamic limit, how might we access this resource? This is the topic of several recent efforts,
for example by scattering \cite{DeChiara}, and is the topic of ongoing investigations. Secondly, with
respect to condensed matter issues, we can ask how these relate to critical phenomena. In this respect
it is intriguing that our results indicate the requirement for certain gapping to observe the survival
of entanglement. Though this and the general connection to entanglement remain large open questions, as
mentioned, with the connection to other thermodynamic quantities, we hope that these connections can be
pushed further and our understanding increased.

\acknowledgments
We are grateful to G. Raggio and Y. Nakata for useful comments and discussions. This work was sponsored by the Asahi Glass Foundation, the JSPS and the European Union (QICS). J.A. acknowledges support of the
Gottlieb Daimler und Karl Benz-Stiftung. V.V. thanks the EPSRC in UK and the European Union for financial support.

%
\end{document}